\documentclass[structabstract]{aa} 

\usepackage[varg]{txfonts}
\usepackage{graphicx}
\begin{document}

\title{An early detection of blue luminescence by neutral PAHs in the direction of the yellow hypergiant HR\,5171A?}

\titlerunning{Transient blue luminescence in the direction of HR5171A?}
\author{A.M. van Genderen\inst{1}
 \and H. Nieuwenhuijzen\inst{2}
  \and A. Lobel\inst{3}} 
\institute{Leiden Observatory, Leiden University, Postbus 9513, 2300RA Leiden, The Netherlands
 \and SRON Laboratory for Space Research, Sorbonnelaan 2, 3584 CA Utrecht, The Netherlands
  \and Royal Observatory of Belgium, Ringlaan 3, 1180 Brussels, Belgium}
 \date{Received.../ Accepted....}

\abstract {} {We re-examined photometry ($VBLUW$, $UBV$, $uvby$) of the yellow hypergiant HR\,5171A  made a few decades ago. In that study no proper explanation could be given for the enigmatic brightness excesses in the $L$ band ($VBLUW$ system, $\lambda_{\rm eff}$\,=\,3838\,\AA). In the present paper, we suggest that this might have been caused by blue luminescence (BL), an emission feature of neutral polycyclic aromatic hydrocarbon molecules (PAHs), discovered in 2004. It is a fact that the highest emission peaks of the BL lie in the $L$ band. Our goals were to investigate other possible causes, and to derive the fluxes of the emission.}
{We  used  two-colour diagrams based on atmosphere models, spectral energy distributions, and different extinctions and extinction laws, depending on the location of the supposed BL source: either in Gum48d on the background or in the envelope of HR\,5171A.} 
{False $L$--excess sources, such as a hot companion,  a nearby star, or  some instrumental effect, could be excluded. Also, emission features from a hot chromosphere are not plausible. The fluxes of the $L$ excess, recorded in the data sets of 1971, 1973, and 1977 varied 
(all in units of $10^{-10}$\,Wm$^{-2}\mu$m$^{-1}$) between 1.4 to 21, depending on the location of the source. A flux near the low side of this range is preferred.  Small brightness excesses in $uv$ ($uvby$ system) were present in 1979, but its connection with BL is doubtful. For the $L$ fluxes we consider the lowest values as more realistic. The uncertainties are 20--30\,\%.  
Similar to other yellow hypergiants, HR\,5171A showed powerful brightness outbursts, particularly in the 1970s. A release of stored H-ionization energy by atmospheric instabilities could create BL emitted by neutral PAHs.} {}

\keywords{Interstellar medium, nebulae: molecules -- Interstellar medium, nebulae: HII regions -- Stars: HR\,5171A -- Stars: hypergiants -- Technique: photometric}

\maketitle

\section{Introduction} 
The fluorescence of small neutral  polycyclic aromatic hydrocarbon molecules (PAHs) in the\object{ Red Rectangle (RR) nebula} in 2003, and in reflection nebulae, was discovered by Vijh et al. (2004, 2005a; 2006, respectively). As this emission is contained in a broad band, somewhere between 3550 and 4400\,\AA,  with most of the peaks between 3700\,\AA, and 4050\,\AA, it was called blue luminescence (BL). Vijh et al. (2005a) found a very close correlation
between the BL distribution and the 3.3$\,\mu$m UIR-band emission. The latter is attributed to the C--H stretch band radiated by neutral PAH molecules.
The central source of the RR nebula is \object{HD\,44179}, a binary, and presumably a post-AGB star of spectral type A ($T_{\rm eff}$\,$\sim$~8000\,K) on
the verge of becoming a planetary nebula. 
To be neutral, PAH molecules should be shielded from direct radiation. In the case of the RR nebula, they probably lie in the
shadow of the optically thick dust ring.  

The present paper deals with a presumed BL effect in the $L$ band ($\lambda$\,=\,3680--4080\,\AA) of the Walraven $VBLUW$ photometric system of the yellow hypergiant
(YHG) HR\,5171A = V766 Cen, reported by van Genderen (1992, Figs.\,1 and 5, hereafter Paper I), made in 1971 and 1973 (and 1977, Sects.\,3.1., 3.3--3.3.1.). Hereafter, these data sets
are abbreviated to sets 71, 73, and 77, respectively. No proper explanation could be offered at the time. Surprisingly, that excess appeared to be absent in the $VBLUW$ time series 1980--1991. 
We also investigated $UBV$ photometry of\object{ HR\,5171B} made in 1971, and $uvby$ photometry of HR\,5171A, made in 1979 and between 1991--1994. The rationale was that the
blue and ultraviolet bands have some overlap with the BL spectrum considering its FWHM~$\sim$~450\,\AA.\footnote{ Geneva seven-colour photometry exists as well, made in the 1970s and 1980s. The catalogue only lists two averages based on an unknown number of observations, and precise dates are lacking (Rufener 1988).} Thus, if the excess was really due to BL, the location could be either Gum48d in the background, or the dense CS matter of HR\,5171A. 

HR\,5171A is thought to be located in the old HII region \object{Gum48d}, known for its gas, dust, and molecular complexes. However, this would implicate that HR\,5171A has a  luminosity that is too high for a hypergiant, with the consequence that its distance should be smaller (Nieuwenhuijzen et al. in prep.).     
Moreover, Karr et al. (2009) concluded that the present energy budget of Gum 48d only needed one ionizing O-type star, i.e the present B-type star HR\,5171B a few million years ago  on the main sequence, just like HR\,5171A roughly at the same time. Thus, this might be another argument that HR\,5171A is no member of 
Gum48d, unless the contribution of the secondary's brightness (unknown; Sect.\,2.) would cancel this inconsistency.

\section{Characteristics of HR\,5171A and the Gum48d nebula}
\object{HR\,5171A\,=\,V766 Cen\,=\,HD\,119796}, in 1971 of spectral type G8Ia$^{+}$ (Humphreys et al. 1971), T$\sim$~5000\,K (Sect.\,3.3.) is an evolved massive star, showing various
types of photometric instabilities. It is suspected to be a contact binary surrounded by a mid-IR nebula (Chesneau et al. 2014). It belongs to the very small group of yellow hypergiants (YHG; de Jager 1980, 1998; de Jager \& van Genderen 1989).
The distance according to Humphreys et al. (1971) is $\sim$~3.6\,kpc, but may be less (Section\,1).
The star still exhibits a high mass-loss rate (Schuster et al. 2006; Schuster 2007; Karr et al. 2009; Chesneau et al. 2014). 
A historical photometric study from 1953 until 1992 was presented in Paper I.  The star showed the signature of a star evolving to the red (which is likely not the case, Nieuwenhuijzen et al. in prep.), in view of the increasing reddening of the colour index $(B-V)_{\rm J}$ ($UBV$ system) between 1960 and 1980 (Paper I; Chesneau et al. 2014). 

Gum48d, containing the photo-dissociation region\object{ RCW\,80} (Schuster et al. 2006; Schuster 2007), is located in the Centaurus Arm. At present, its ionization is sustained by the near-by blue star HR\,5171B (Schuster 2007), a B0Ibp star (Humphreys et al. 1971), at a distance of $\sim$~10$\arcsec$. Most of the UIR bands of Gum48d contain emission features of PAHs (Schuster et al. 2006; Schuster 2007; Karr et al. 2009). The 3.4\,$\mu$m C--H stretch band emission from the WISE archive, which is the best signature for neutral
PAHs (Bakes et al. 2004), turned out to be saturated at a flux of 95\,Jy (Cutri et al. 2012, see also in Chesneau et al. 2014, the Appendix Table D.1).
Thus, based on these observations, the nebula is a serious candidate to be the source for the BL. On the other hand, Chesneau et al. (2014), believe that HR\,5171A is responsible for the high far- and mid-IR excess (see their Table\,D.1.) as the instruments used have apertures not larger than a few arcsec, centred on HR\,5171A. 
However, with respect to temperature, HR\,5171A is too cool ($\sim$~5000\,K) to produce high energetic photons (3.5--5\,eV; Witt 2013, priv.comm.) to create PAH molecules in the highest energetic state. The presence of atmospheric instabilities, during which a huge amount of energy is released like in some other YHG (Lobel et al. 2003; Nieuwenhuijzen et al. 2012), is suspected. These processes could be well sufficient to produce these kinds of photons (Sect.\,4). 

HR\,5171A shows a
prominent 10\,$\mu$m silicate feature (Humphreys et al. 1971) and a strong Na\,I 2.2\,$\mu$m emission (Chesneau et al. 2014), also detected in other
massive evolved stars like the YHG (Oudmaijer \& de Wit 2013). This could point to the presence of a pseudo-photosphere in an optically thick wind, shielding
the neutral sodium from ionization by direct starlight (Oudmaijer \& de Wit 2013). These authors also claim that the NaI emission in other YHG originates in
a region inside the dust condensation radius.  
Considering the  controversy about the precise origin of the $L$ excess: the CS material of HR\,5171A and Gum48d, we compute the fluxes for these two alternatives. A possible instrumental effect responsible for a false emission in the $L$ channel during the 1970s is also  discussed (Sect.\,3.1).
We also discuss two emission features radiated by a hot chromosphere, often a feature exhibited by cool stars, which under abnormal hot and active conditions also radiate in the wavelength area of the $L$ channel (Sect.\,3.2.). 

\section{The photometric observations}
\subsection{The $VBLUW$ photometry. Other possible causes to explain the $L$--excess}
A total of seven observations, made between 1971 and 1977, showing the strong brightness excess in the $L$ band ($\lambda_{\rm eff}$\,=\,3838\,\AA), were obtained with the 90-cm Dutch telescope, which was then still located at the Leiden
Southern Station in South Africa, Broederstroom (Hartebeestpoortdam). At the time, the telescope was equipped with the same simultaneous five-colour photometer of Walraven as
between 1980--1991 (no $L$ excess) after the telescope was moved to the ESO in 1979, La Silla,  Chile (Lub \& Pel 1977; de Ruiter \& Lub 1986; Pel \& Lub 2007). It is customary to express the brightness and colour indici of the $VBLUW$ system in log intensity scale.
The observing and calibration procedures were similar for all data sets (Paper I).
However, the aperture was 21$\arcsec$5 and/or 16$\arcsec$5 in the seventies (original telescope outputs are not available anymore), and 11$\arcsec$6 in the eighties. In both cases, the near-by blue star
HR\,5171B, in the visual 3$\fm$5 fainter than HR\,5171A, could be satisfactorily kept out by decentring HR\,5171A with 3$\arcsec$--5$\arcsec$.

\begin{figure}
\resizebox{\hsize}{!}{\includegraphics[trim = 0mm 0mm 0mm 0mm, clip]{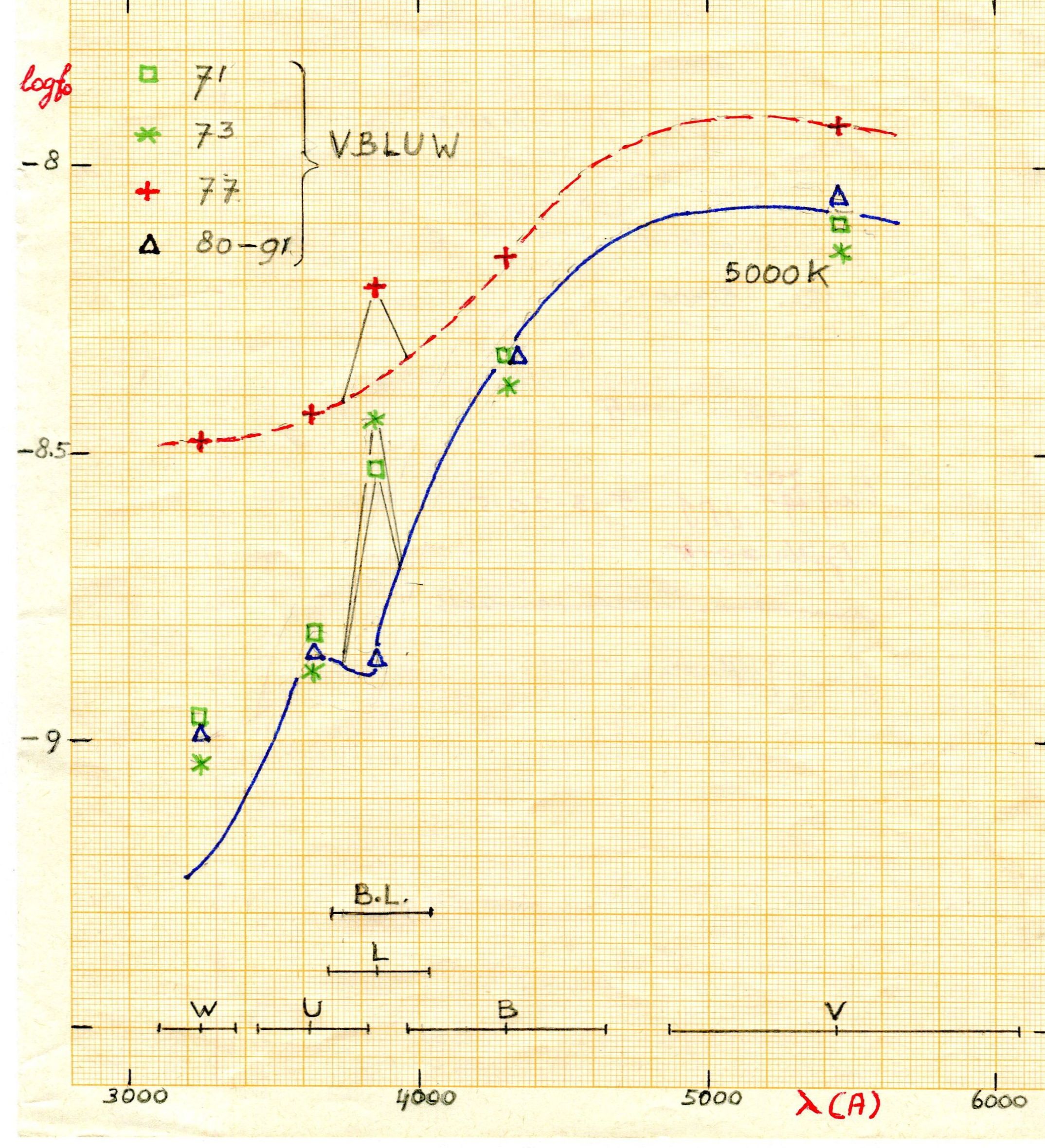}}
\caption{The fluxes log\,f$_{0}$ for the $VBLUW$ data sets as a function of the wavelength. The theoretical SED for a star with log\,g\,=\,0, and T\,=\,5000\,K (blue curve) is fitted to the 71, 73 and 80--91 fluxes. The red dashed curve fitting the 77 set fluxes (apart from the $L$ flux as it includes the suspected BL) does not agree with any model. At the bottom, the filter coverage for the $VBLUW$ bands, and the BL spectrum for only the highest peak. The $L$ excesses are indicated with triangular peaks.}
\label{fluxes}
\end{figure}

Two observations were obtained in 1971 (Pel 1976), two observations in 1973 (Paper I) and three in 1977 (Pel 2013, priv. comm.). The time intervals between the individual observations in a set were a few days up to a few weeks. All photometric data are given in the 1980 system (Paper I).
Those made in 1971 and 1973, on both sides of maximum no.\,3 (see Fig.\,1 of Paper I) are marked '$L$ excess'. 
The 1977 observations are located in the rising branch to maximum no.\,7, very near the top (not plotted in Fig.\,1 of Paper I, as we obtained them from Pel only recently. The star was brighter and bluer than usual because of intrinsic changes (Sect.\,3.3.).

The amount of $L$ excess is based on the position in the theoretical two-colour diagram $V-B$/$B-L$ (empirical version in Fig.\,5 in Paper I) based on Kurucz-atmospheres (Castelli \& Kurucz 2003; Castelli 2014).  The excesses in $B-L$ represent the colour index difference between the observed and expected position (at roughly constant 
$V-B$), if there was no excess in $L$.

Table\,1 lists the observed photometric parameters for the three data sets of the 70s and the median values of the 80--91 set. The parameters $V_{\rm J}$ and $(B-V)_{\rm J}$ were transformed from $V$ and $V-B$ by formulae of Pel 1987).
The last column lists the $L$ excesses, labelled $L$Ex.

The $L$ excesses in the early 1970s, which is very obvious in the empirical two-colour diagram $V-B$/$B-L$, could not be due to any stray light from the blue nearby star HR\,5171B; whatever the size of the diaphragm, as HR\,5171A was always decentred in such a way that the blue star's distance to the edge was enlarged. Moreover, HR\,5171A would have shown progressively increasing excesses in $U$ and $W$, if this was done insufficiently. The suggested presence of a mysterious blue star C in the diaphragm, at a 0$\arcsec$15 distance, nearly as bright as HR\,5171A, \footnote {(Dommanget \& Nys 1994; Dommanget 2002; Mason et al. 2001 (WDS13472-6235); Mason \& Hartkopf 2006; ESA 1997 (HIP 67261); van Leeuwen 2012, priv.comm.). However, its existence is still a matter of debate: HST images do not show this component (Schuster \& Humphreys 2014, priv.comm. On the other hand, the analysis of three years of Hipparcos photometry, revealed two 4$^{\rm d}$5 oscillations with a light amplitude of 0\fm05 (Eyer 1998). Thus, a variable star should be inside the aperture used by the Hipparcos satellite.} is unlikely. Otherwise, the $U$ and $W$ fluxes should have shown even larger excesses. Moreover, the 71/73 $B-U$ and $U-W$ indices  (in 1977 the star brightened intrinsically, Sect.\,3.3.) were very similar to those of 80/91 lacking any $L$ excess! 

Another possible explanation of the false $L$ excesses, with the observed intensity, is if the 'field of view' of the photomultiplier of the $L$ channel was accidentally shifted by $\sim$~5$\arcsec$, (with respect to the other channels) in such a manner that HR\,5171B was shifted up to, or on the edge of the diaphragm. This scenario is considered  almost impossible from a technical point of view (Pel 2014, priv.comm.). Moreover, if this had been the case, it would mean that hundreds of program stars and tens of thousands of $L$ measurements of comparison and standard stars during the seventies, were badly centred by about 5$\arcsec$. The much less favourable seeing conditions at the location of the Southern Station, compared to the seeing at the ESO, would have entailed a much larger scatter for $B-L$ than for the three other colour indices. A quality check of, e.g. the 150 and 100 complete light and colour curves of Population\,I Cepheids (Pel 1976) and RR\,Lyrae  stars (Lub 1977), respectively, proves that this was not the case.

\begin{table*}
\caption{The observed photometric parameters of the four data sets and the brightness excesses in the $L$-band. The $L$ (the total brightness in the $L$-band) and $L$Ex. are means of the data set. Only
$V_{\rm J}$ and $(B-V)_{\rm J}$ are in magnitudes, the remainder is given in logarithmic scale}
\label{obs}
\centering
\begin{tabular}{|c c | c c | c c c c c c c|} 
\hline\hline
JD--244 0000 & set & $V_{\rm J}$ & $(B-V)_{\rm J}$ & $V$ & $V-B$ & $B-L$ & $B-U$ & $U-W$ & $L$ & $L$Ex. \\
\hline
1164.5       &  71 & 6.80 & 2.30 & -0.004 & 1.124 & 0.528 & 1.010 & 0.50 & -1.660 & 0.25 \\
1169.5       &       & 6.75 & 2.31 &  0.017 & 1.126 & 0.555 & 1.001 & 0.60 \\ 

1758.6       &  73 & 6.88 & 2.34 & -0.034 & 1.140 & 0.395 & 1.042 & 0.60 & -1.572 & 0.37 \\
1787.5       &       & 6.90 & 2.30 & -0.042 & 1.122 & 0.410 & 0.987 & 0.54 \\

3248.5       &  77 & 6.41 & 2.31 &  0.154 & 1.125 & 0.394 & 0.813 & 0.39 & -1.343 & 0.18 \\
3252.5       &       & 6.35 & 2.29 &  0.178 & 1.121 & 0.395 & 0.781 & 0.48 \\
3269.5       &       & 6.32 & 2.27 &  0.191 & 1.109 & 0.407 & 0.796 & 0.48  \\

4312--8315   & 80--91 & 6.60 & 2.55 & 0.060 & 1.23  & 0.83  & 0.98 & 0.54 & -2.00 & 0 \\ 
\hline
\end{tabular}
\end{table*}

\subsection{Other emission sources in the wavelength area of the $L$ band}
Two emission sources, which can give rise to radiation in the wavelength area of the $L$ band, are related to a hot chromosphere. Cool stars like the twin sister of HR5171\,A, $\rho$~Cas, at times showed the signature of a hot chromosphere. One presumes that this is caused by the dissipation of mechanical energy delivered by the different types of pulsations and occasional shell ejections (Sargent 1961; Zsoldos \& Percy 1991; Percy \& Kolin 2000; Lobel et al. 2003). This object showed a variable Balmer continuum radiation starting at $\sim$~4200\,\AA, up to shorter wavelengths, in the absolute energy distribution for two nights in 1970 and two in 1974 (Joshi \& Rautela 1978). The variable amplitude of this radiation at the location of the $L$ band was $\sim$~0\fm3. As a second emission source, we have to mention the CaII H and K resonance lines, located in the red wing of the $L$ band. Normally these resonance lines are in absorption, but in a very heated chromosphere they appear in emission, as in flare stars. There are convincing reasons to reject a hot chromosphere as the source of the $L$  excesses in the 1970s (Sects.\,3.3, 4.).

\subsection{On the location of the BL. The extinctions. The spectral energy distribution (SED) for the $VBLUW$ data sets}

An important issue is the location of the supposed BL. There are two possibilities. Scenario (1) in Gum48d, or (2) in the extended envelope of HR\,5171A. Depending on the location, different corrections for extinction (interstellar IS and/or circumstellar CS) should be applied to the observed $L$ fluxes. 

Scenario (1) In this case the reddening of HR\,5171B has been used: E$(B-V)_{\rm J}$\,=\,1.00\,$\pm$0.10, a rounded-off average of the reddenings based on the photometry by Humphreys et al. (1971), Dean (1980) and Paper I. Using the extinction law for the diffuse IS extinction R$_{\rm J}$\,=\,3.2 (Cardelli et al. 1989), A$_{Vj}$\,=\,3.20\,$\pm$\,0.32. The equivalent values in the Walraven $VBLUW$ system are: E($V-B$)\,=\,0.43\,$\pm$\,0.04 (log scale, transformation formula by Pel 1987), and  R$_{\rm W}$\,=\,3.2, in log scale: A$_{V}$\,=\,1.38\,$\pm$\,015, and in the $L$ band: A$_{L}$\,=\,1.97\,$\pm$\,0.15 (Steenman \& Th\'e, 1989). The calibration constant is given by de Ruiter \& Lub (1986). 

Scenario (2) This scenario is  of importance for the observed, non-stellar $L$ excesses and for the stellar SED. The total reddening for the four data sets was determined with the reddening lines in the two-colour diagrams defined by Pedicelli et al. (2008), and the position of the 80-91 set as reference. This is because it was free of any $L$ excess and if dereddened, its position in the $V-B$/$B-L$ and $B-U$/$U-W$ diagrams moves to T$\sim$~5000\,K, adopting log\,g\,=\,0. This is in agreement with spectral analyses (Humphreys et al. 1971; Warren 1973; Chesneau et al. 2014; Lobel 2014, priv.comm.; Oudmaijer 2014, priv.comm.). 

The reddening (IS and CS) in log scale was E($V-B$)\,=\,0.63\,$\pm$\,0.04, corresponding with
E$(B-V)_{\rm J}$\,=\,1.41\,$\pm$\,0.10 (in magnitudes, transformation formula from Pel 1987). It appeared that for luminous G-type stars the extinction law is 
R$_{\rm J}$\,=\,3.5--3.6 (Humphreys 2014, priv.comm.) because of their extended envelopes with gas, dust, and molecular clouds. To find the R values for the three photometric systems Johnson $UBV$, Str\"omgren $uvby$ , and Walraven $VBLUW$, we used observations (almost) simultaneously made in the three systems. We found R$_{\rm J}$\,=\,3.6, R$_{\rm S}$\,=\,4.7, and R$_{\rm W}$\,=\,3.3 for which we got a satisfactory match of the three SEDs.\footnote{This matching eliminated small reddening differences, which are evident from the $V-B$ values between the 70s and 80s; see Table\,1. We emphasize that these R values are only applicable to this star. The reason is that the cause of the peculiar and long-lasting stellar reddening is not well understood.}
Hence, the extinctions are A$_{Vj}$\,=\,5.08\,$\pm$0.30 (in mag) and A$_{L}$\,=\,2.96\,$\pm$0.15 (in log scale). 
The extinction-free stellar fluxes log\,f$_{0}$ resulted from the addition of the calibration constants (de Ruiter \& Lub 1986) and the  extinction-free brightnesses. They are listed in Table\,2.  

\begin{table}
\caption{The  extinction-free brightnesses log\,I$_{0}$ and fluxes log\,f$_{0}$ for the five channels of the four data sets according to scenario (2)}
\label{Fluxes}
\centering
\begin{tabular}{|c c|c c|}
\hline\hline
Set & band & log\,I$_{0}$ & log\,f$_{0}$ \\
\hline
71 & V & 2.086 & -8.086 \\
   & B & 1.592 & -8.318 \\
   & L & 1.300 & -8.518 \\
   & U & 0.986 & -8.807 \\
   & W & 0.716 & -8.957 \\
\hline
73 & V & 2.042 & -8.130 \\
   & B & 1.541 & -8.360 \\
   & L & 1.388 & -8.430 \\
   & U & 0.926 & -8.867 \\
   & W & 0.636 & -9.037 \\
\hline
77 & V & 2.254 & -7.918 \\
   & B & 1.766 & -8.144 \\
   & L & 1.617 & -8.201 \\
   & U & 1.369 & -8.424 \\
   & W & 1.199 & -8.474 \\
\hline
80--91 & V & 2.140 & -8.032 \\
       & B & 1.593 & -8.317 \\
       & L & 0.960 & -8.858 \\
       & U & 0.960 & -8.833 \\
       & W & 0.710 & -8.963 \\
\hline
\end{tabular}
\end{table}

Fig.\,1 shows the  SEDs of HR\,5171A  for scenario (2) only;\ the green and blue symbols denote the 71, 73, and 80--91 sets, respectively, and the red symbols denote the 77 set. It is obvious that the brightness differences between the sets 71, 73 and 80--91 were small, therefore, they could be used as one homogeneous set. It appears that a satisfactory match, even with the small positive Balmer jump, is obtained for $V$, $B$, ($L$ only for the 80-91 set), and 
$U$ with the model SED for 5000\,K, adopting log\,g\,=\,0, and v$_{\rm turb}$\,=\,2\,kms$^{-1}$ , using Kurucz models (Castelli \& Kurucz 2003; Castelli 2014).  The $W$ flux is too high by about 0.2 log scale ($\sim$~0\fm5) for all three sets, thus, possibly caused by chromospheric radiation. Obviously, these $W$ fluxes represented very weak Balmer continuum excesses, otherwise $U$ should show excesses as well, which should then be greater than those in $L$. Therefore, these $L$ excesses should be due to some other source. The $L$ excesses are indicated with the two triangles. 

The SED of the 77 set is a different story: no match with any hotter theoretical SED is possible. The slope of the Paschen continuum (from $V$ to $B$), is slightly steeper than for the other sets. Thus, a possible photospheric temperature rise is small  (by a few hundred degrees), but the extreme high excess in the ultraviolet is stunning, with on top the $L$ excess. The pulsational activity and/or outbursts of ionization energy, represented by an excess of optical light (see Fig.\,1 in Paper I), heated the chromosphere presumably to such a high level that it resulted in an excessive Balmer continuum emission of up to more than 1$^{\rm m}$ in $U$ and $ W$ (Sect.\, 4.). It should be emphasized that the observed $L$ brightnesses (71, 73, and 77) contain a stellar and a non-stellar contribution. The stellar contribution was likely subject to a much higher reddening than the non-stellar one (Sect.\,3.3.1.).
  
\subsubsection {The $L$ excess fluxes}    

\begin{table*}
\caption{The computation of the fluxes f(E$_{L0}$) (in $10^{-10}$\,Wm$^{-2}\mu$m$^{-1}$) of the BL, based on the brightness excesses in the $L$ band after correction for interstellar extinction and assuming that: (1) the source lies in Gum48d ; column f(E$_{L0})$(1), and (2) that the source lies in the CS envelope of HR\,5171A; column f(E$_{L0})$(2). See Sect.\,3.3.1.}
\label{fluxL}
\centering
\begin{tabular}{|cccc|c|c|}
\hline\hline  
set & $L$ & $L$- $L$Ex.  & E$_{L}$  & f(E$_{L0})$(1) & f(E$_{L0})$(2) \\
\hline
71    & -1.660 & -1.910 & -2.019 & 1.37 & 14.7 \\
73    & -1.572 & -1.942 & -1.813 & 2.20 & 21.3 \\
77    & -1.343 & -1.523 & -1.812 & 2.21 & 21.4 \\   
\hline
\end{tabular}
\end{table*}

In  Table 3, we list the date (set); the brightness of HR\,5171A in the $L$ band (taken from Table\,1),
 including the excess $L$Ex; the corrected brightness of HR\,5171A in $L$ without the excess
($L$-$L$Ex); the brightness of the excess (E$_{L}$),  which is the difference of the
anti-logs of $L$ and $L$-$L$Ex., after transforming it into a log scale; and the unreddened fluxes, f(E$_{L0}),$ from scenario (1), in $10^{-10}$\,Wm$^{-2}\mu$m$^{-1}$. The uncertainty 
is $\pm$\,0.5, and, thus between 30--20\%, and mainly due to the 0\fm10 uncertainty in the reddening. As the reddening differences between the three data sets were small (Sect.\,3.3.), the difference in the fluxes of 1971 and 1973/1977 should be considered  real and suggests that the presumed BL was variable. 
The diameter of the apertures used were $21\arcsec$5, and/or $16\arcsec$5 (1971--1977, Sect.\,3.1.), and $11\arcsec6$ (1980--1991). These apertures are 
equivalent with an area with a diameter of 0.4, 0.3 and 0.2\,pc in Gum48d, respectively, assuming a distance of 3.6\,kpc. 
The effect of the Gum48d's radiation travelling through HR\,5171A's envelope, with its much higher reddening (see scenario [2]), can be neglected because its diameter is about $3\arcsec4$ (Chesneau et al. 2014), much smaller than the size of the apertures used.  

Scenario (2): The dereddened  fluxes f(E$_{L0})$, are listed in the sixth column.
It appears that the fluxes are now larger than for scenario (1) by a factor of ten (due to the higher reddening) as well as the uncertainty. Thus, 14.7 in 1971 and 21.4\,$\pm$\,5 in 1973/1977. 

Some doubt about the reliability of these high values, almost of the order of the stellar $L$ flux, is appropriate. However, it is likely that the supposed BL source did not suffer from the same high extinction as the central star HR\,5171A because its location could have been near the border of the envelope (e.g. facing the Earth), where the extinction is much lower. Then, the lower reddening used in scenario (1) would be much more appropriate, resulting in fluxes about ten times smaller. 
The envelope of HR\,5171A may be more or less similar to the envelope surrounding the AGB star\object{ IRC\,+10216 (CW Leo)}, a Mira Ceta variable), modelled by McElroy et al. (2013), Li et al. (2014) and Li (2015). That model has an A$_{\rm v}$\,$\sim$\,14\fm0 in the centre (for HR\,5171A: $\sim$~5$^{\rm m}$), decreasing exponentially outward and being almost negligible halfway to the edge. Thus, by virtue of this model, it is much more realistic to suppose that the fluxes for the $L$ excesses should have been near the low side of the total range 1.4 to 21. A dust model of HR\,5171A by Lobel (2015, priv. comm.) will be released soon. 

Thus, then the remarkable size of the $L$ excesses, as evident in Table\,3, and in the empirical $V-B$/$B-U$ diagram in Fig.\,5 in Paper\,I, being almost of the order of the stellar flux, can be understood (see our comment on the size of the 1977 flux below, and in Sect.\,4). Moreover, the star was, and still is, intrinsically very red.

It is evident that the flux measured in the 77 set contains the excess of the unknown source and those radiated by the hot chromosphere. Thus, the value listed in Table\,3 represents the total excess flux of different sources.
   
\subsection{The search for any emission in the $U_{\rm J}$ band (Johnson $UBV$ system), and in the $uv$ bands ($uvby$ system)}

 Any $uv$ observation of HR\,5171A, made after the 1971--1977 time interval, and one $U_{\rm J}$ observation of its close neighbour HR\,5171B, made in 1971 (by Humphreys 2014, priv.comm.), has been carefully inspected and analysed. Despite their partial overlap with the BL spectrum no excess in $U_{\rm J}$ was found, and in $uv$ only a small amount of excesses, which we do not consider  conclusive. Therefore, they are only briefly discussed.

As the $U_{\rm J}$ magnitude of HR\,5171B in 1971 was not affected by any measurable emission, we conclude that either the contribution to the $L$ brightness of set 71 originated accidentally behind HR\,5171A in a small area of Gum48d, or its origin was the CS material of HR\,5171A.     

 Olsen (1983) obtained seven observations in the Str\"omgren $uvby$ system in the period 1979, February 26 and October 6, with the Danish 50-cm
reflector at the ESO in La Silla, Chile, equipped with the simultaneous $uvby$ photometer (Gr\mbox{\o}nberg et al. 1976; Gr\mbox{\o}nberg \& Olsen 1976). The aperture was $30\arcsec$ and included HR\,5171B, for which we had to correct before inspecting the $v$ magnitude of HR\,5171A
for any brightness excess.
The mean  errors in the average photometric parameters appear to be very small (Olsen 1983). 

As no $uvby$ photometry has ever been made of HR\,5171B, we had to use models (solar abundancy) for early B-type stars: those of 
Lester et al. (1986) and of Castelli \& Kurucz (2003)/Castelli (2014). We adopted T$_{\rm eff}$\,=\,26\,500\,K, log\,g\,=\,3 and 3.5, and a $y$ magnitude almost equal
to the observed $V_{\rm J}$\,=\,10.03 (Paper I). The $V_{\rm J}$ ($\sim$~$y$) above was dereddened using E$(B-V)_{\rm J}$\,=\,1.00\,$\pm$\,0.10 (Sect.\,3.2), and R$_{\rm J}$\,=\,3.2. For the amount of extinction in the $uvby$ channels, we consulted the tables of Steenman \& Th\'e (1989).    

\begin{table*}
\caption{The comparison of the observed 1979 $uvby$ magnitudes of HR\,5171A (corrected for the light of HR\,5171B) on the third line with the 1991 observed $uvby$ magnitudes on the fourth line.  The mean reddened model of HR\,5171B is on the second line.}
\label{str}
\centering
\begin{tabular}{|l|cccc|}
\hline\hline
set/type/object                                    & $y$    & $b$   & $v$   & $u$  \\
\hline
set 79, observed, HR\,5171A\,+\,HR\,5171B                    &  6.78  &  8.45 & 10.16 & 11.45 \\
mean reddened model HR\,5171B                      & 10.03  & 10.67 & 11.20 & 11.86 \\
\hline
set 79, "observed", HR\,5171A (corr. for star HR\,5171B)  &  6.84 &  8.60 & 10.68 & 12.71 \\
set 91, observed, HR\,5171A                             &  6.74 &  8.51 & 10.89 & 13.24 \\
\hline
\end{tabular}
\end{table*}

Table\,4 lists Olsen's $uvby$ observations of both stars together (first line), the mean of the two models for the
blue star (second line), and then in the third line the "observed" models of HR\,5171A only. Thus, the magnitudes on the second line subtracted from the magnitudes on the first line. The estimated uncertainties in the models are smaller than 0\fm1, with the
exception of the $u$ magnitude, which is $\pm$\,0\fm2. The fourth line in Table\,4 lists the mean of three $uvby$ observations made in 1991 (see below). 

Str\"omgren $uvby$ observations were also made by the LTPV Group of Sterken (1983) from 1991 until 1994 with the 50-cm ESO telescope in La Silla, Chile (Manfroid et al. 1991; Sterken et al. 1993). 
The very first three $uvby$ observations by the LTPV Group, coincided with the last six observations of the 80--91 $VBLUW$ data set in maximum no.\,17 (Fig.\,1 in Paper I; van Leeuwen et al. 1998). We used these three 1991 $uvby$ observations to compare them with the 1979
observations of Olsen (listed in Table\,4, third line). Although the aperture used by the LTPV group was 20$\arcsec$, star HR\,5171B could generally be excluded by decentring HR\,5171A with $\sim$~5$\arcsec$. 

\begin{figure}
\resizebox{\hsize}{!}{\includegraphics[trim = 0mm 0mm 0mm 0mm, clip]{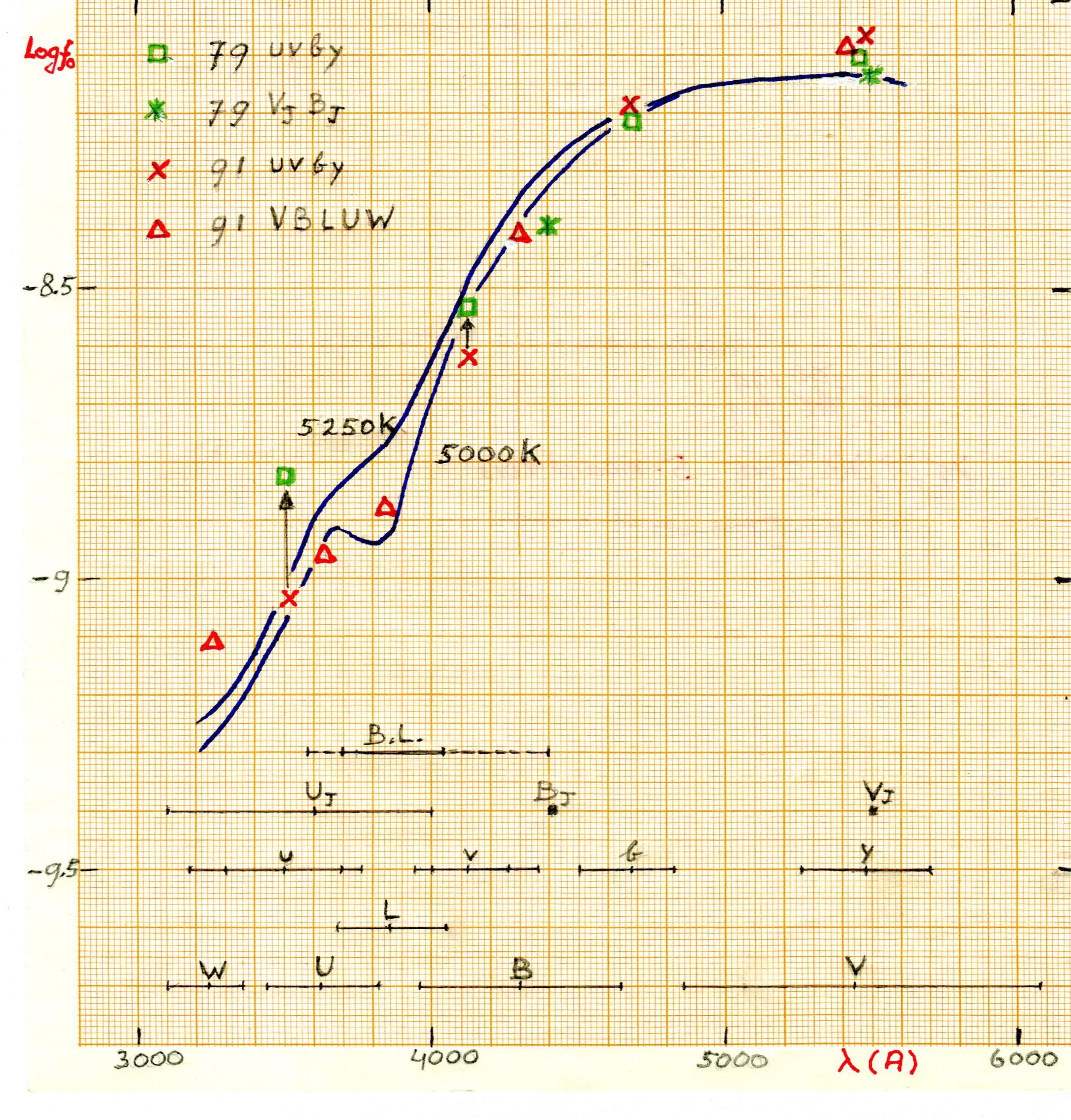}}
\caption{Log\,f$_{0}$, for observations made almost simultaneously in two different photometric systems (green and red symbols) as a function of $\lambda$ with the indication of the $\lambda$ coverage. $UjBjVj$ for Johnson, $uvby$ for Str\"omgren, $VBLUW$ for Walraven. The blue curves are SEDs for 5000\,K and 5250\,K models. Vertical arrows: see the text in Sect.\,3.5.}
\label{logf$_{0}$}
\end{figure}

\subsection {The spectral energy distribution (SED) for the 1979 and 1991 data sets in three photometric systems.}

 We constructed an SED (Fig.\,2) simultaneously using obtained $B_{\rm J}V_{\rm J}$ ($UBV$ system) photometry by Dean (1980); see Fig.\,1 in Paper I, together with the simultaneously observed $uvby$ and $VBLUW$ data from the 1991 sets. The band widths are indicated at the bottom with horizontal line pieces. The central mark indicates $\lambda_{\rm eff}$. For $v$ and $u,$ two widths are indicated: the smallest  was used by the LTPV group (Sterken 1983), and the largest was used by Olsen (1983). The $B_{\rm J}$ and $V_{\rm J}$ bands are only represented by their $\lambda_{\rm eff}$. The band width (FWHM) of the BL for the RR nebula is indicated (Vijh et al. 2004, 2005b, 2006). The full line is the portion containing the highest emission peaks.
The same reddening was used as in Sect.\,3.3. (scenario\,2), as well as the extinction laws for the three photometric systems, and the extinctions for each pass band. Calibration constants are from Cox (1999), for 
$UBV$; Helt et al. (1991); for $ubvy$, and de Ruiter \& Lub (1986), for $VBLUW$.

\begin{table}
\caption{The reddening free fluxes of HR\,5171A (Wm$^{-2}\mu$m$^{-1}$) measured in three different photometric systems. Each pair of sets was simultaneously observed. The last column lists the mean visual magnitude. In the last  data set from 1991, the $V_{\rm J}$ magnitude represent the computed equivalent value of $V$.}  
\label{sed}
\centering
\begin{tabular}{|c|ccc||c|ccc|}
\hline\hline
set & band & log\,f$_{0}$ & $V_{\rm J}$  & set & band & log\,f$_{0}$ & $V_{\rm J}$ \\
\hline
79 & $y$ & -8.093 & 6.84 & 91 & $y$ & -8.053 & 6.74 \\
   & $b$ & -8.200 & 8.60  &   & $b$ & -8.166 & 8.51 \\
   & $v$ & -8.526 & 10.68 &   & $v$ & -8.609 & 10.89 \\

   & $u$ & -8.813 & 12.71 &   & $u$ & -9.025 & 13.24 \\
\hline
79 & $V_{\rm J}$ & -8.126 & 6.83 & 91 & $V$ & -8.058 & 0.044 \\
   & $B_{\rm J}$ & -8.386 & 9.47 &    & $V_{\rm J}$  & & 6.67 \\
   &             &        &      &    & $B$ & -8.393 & -1.186 \\
   &             &        &      &    & $L$ & -8.867 & -2.021 \\
   &             &        &      &    & $U$ & -8.945 & -2.232 \\
   &             &        &      &    & $W$ & -9.092 & -2.806 \\
\hline 
\end{tabular}
\end{table}

 The computed fluxes log\,f$_{0}$ are listed in Table\,5.
The obtained SEDs of the simultaneous sets 79 $uvby$ and $B_{\rm J}V_{\rm J}$, match very satisfactorily. They also match well the 91 sets  $uvby$ and $VBLUW$. Fig.\,2 shows  brightness excess by about 0.1-0.2 (see arrow) in the $u$ band in log\,f$_{0}$ with respect to the combined SED. However, considering the errors,
we are very hesitant to consider it as due to BL. Errors in the models and physical differences of the star between 1979 and 1991 are possible, but some contribution by Balmer continuum radiation is possible as well.
The excess in $W$: $\sim$~0\fm3, represents a weak Balmer continuum emission, but weaker than during the 1970s and 1980s (Fig.\,1: $\sim$~0\fm5, Sect.\,3.3.

\section{Discussion and conclusions}
It is quite well possible that the brightness excesses in the $L$ band ($VBLUW$ system) detected in the photometry of the yellow hypergiant HR\,5171A made in the 1970s were due to blue luminescence (BL) that was discovered in the RR nebula by Vijh et al. (2004; 2005a,b; 2006). Our suspicion is possible in view of the complete overlap of its spectrum, including the highest peaks, with the $L$ band.
We first considered various possibilities, which might have created false brightness excesses in the $L$ band, but they turned out to be implausible. 
The same can be said about the following emission sources previously mentioned in Sect.\,3.2.: the Balmer continuum emission due a hot chromosphere and the CaII H and K resonance lines (in the $L$ band), which only appear in emission in extreme circumstances, when the chromosphere is extremely heated. However, in the first case one expects much higher brightness excesses in $W$ and $U$ than in $L$ (see Fig.\,1 in Joshi \& Rautela 1978, showing an increasing Balmer continuum emission from $\lambda$~4200\,\AA, to shorter wavelengths in the SED of \object{$\rho$~Cas}). It is obvious in our Figs.\,1 and 2 that apart from the 77 set observed during a large amplitude pulsation showing high Balmer continuum excesses in $W$ and $U$, the 71 and 73 sets show no excesses in $U$ and the excesses in $W$ are smaller than in $L$. Note that in the 1980s there was no $L$ excess at all other than a $W$ excess, presumably due to the presence of a chromosphere. We conclude that the $L$ excesses in 1971 and 1973 were not caused by Balmer continuum emission. Obviously, the $L$ excess of the 77 set in Fig.\,1 contains the contribution from the supposed BL source, but also should include Balmer continuum emission, and perhaps also some emission from the CaII doublet. Therefore, the $L$ flux listed in Table\,3 is too high for the supposed BL alone, indicating that this emission was declining. After all, in 1979, no excess in that wavelength area of the $uvby$ system was recorded, apart from some excess in the $u$ band, presumably from uncertainties in the atmosphere models and/or from some Balmer continuum radiation (Fig.\,2).

The unreddened $L$ fluxes, depending on the location of the BL source, in units of $10^{-10}$\,Wm$^{-2}\mu$m$^{-1}$, varied between 1.4 and 2.2 (71 and 73 sets), when located in Gum48d (scenario 1). If the unreddened $L$ fluxes were located in the dense centre of the CS envelope of HR\,5171A (scenario 2) then the values would be a factor of 10 higher, i.e. 15--21. However, these values were much lower, approaching the lower values that we prefer, if they were located near the outskirts of the envelope.    

No significant trace of an excess due to BL was detected in other bands apart from the $L$ band. The distribution of the highest BL peaks in the direction of HR\,5171A occurred in the $L$ band, which does not contradict the possibility that BL was the cause.

As mentioned above, no $L$ excess was present in the 1980--1991 $VBLUW$ time series, while the colour indices $B-U$ and $U-W$ were almost the same as in the early seventies (1971/1973), which is in contrast to the $B-L$ colour index. The short duration of the supposed BL emission could mean that we were dealing with a small-scale source. The dynamical circumstances of the CS matter surrounding HR\,5171A would then be a good candidate, as shielded clouds containing neutral PAH molecules can quickly be exposed to highly energetic photons. On the other hand, one may wonder whether this cool star ({$\sim$~5000\,K}) was able to produce high energetic photons at all to excite the neutral PAH molecules to the upper electronic state. As for this requirement, Gum48d, with its hot exciting star HR\,5171B, was by far the best candidate. 
However, it is very plausible that these photons were produced by HR\,5171A itself.
The star showed a number of relative short-lasting 'light outbursts' during the 1970s, being blue near maximum light (Fig.\,1 in Paper I; see SED of maximum no.\,7 in Fig.\,1 of the present paper). Amplitudes were $\sim$~0\fm5 in $V$, with a timescale of $\sim$~1--2\,yr.
Similar type of light curve features were exhibited by the YHG $\rho$\,Cas (Lobel et al. 2003) and \object{HR\,8752} (Nieuwenhuijzen et al. 2012).
These authors concluded that they showed at times strong atmospheric ionization driven instabilities. As a result, explosive 'flash' outbursts occurred by the release of stored ionization energy due to H recombination as the temperature of the atmosphere declined. These processes might have delivered the necessary high energetic photons (3.5--5\,eV, Witt 2013, priv.comm.).

\begin{acknowledgements}
We are very grateful to a number of colleagues: Prof. Xander Tielens for his support, who drew our attention to the papers of Vijh et al. (2004, 2005a); Prof. Jan Willem Pel for offering his 1977 $VBLUW$ photometry; and the following colleagues for many fruitful discussions, help, and invaluable advices: Dr. Floor van Leeuwen, Prof. Jan Willem Pel, Prof. Adolf Witt, Dr. Jan Lub,  Prof. Roberta Humphreys, and Dr. Ren\'e Oudmaijer, and especially to Dr. Erik-Heyn Olsen, who spent a lot of time  studying his log books, making it easy for us to establish the date of the seven $uvby$ observations. We took much advantage of his and Dr. Chris Sterken's expertise on the Str\"omgren photometric system. 
A.L. acknowledges partial financial support by the Belgian Federal Science Policy Office under contract No. BR/143/A2/BRASS and in connection with the ESA PRODEX programmes 'Gaia-DPAC QSOs'  and 'Binaries, extreme stars and solar system objects' (contract C90290).
We are grateful to the referee for the invaluable comments and suggestions.      
\end{acknowledgements}

\listofobjects

\end{document}